\documentstyle[12pt]{article}

\renewcommand{\d}{{\mathrm d}}

\begin{document}

\title{Active Gravitational Mass and The Invariant Characterization
      of Reissner-Nordstr\"om Spacetime}
\author{ L. Herrera$^1$\thanks{e-mail: laherrera@telcel.net.ve}, N. O. Santos$^{2,3}$\thanks
{e-mail: nos@cbpf.br} and J. E. F. Skea$^4$\thanks{e-mail: jimsk@dft.if.uerj.br}\\
\small{$^1$Escuela de F\'{\i}sica, Facultad de Ciencias,} \\
\small{Universidad Central de Venezuela, Caracas, Venezuela.}\\
\small{$^2$Laborat\'orio Nacional de Computa\,{c}\~ao Cient\'{\i}fica,}\\
\small{25651-070 Petr\'opolis RJ, Brazil.}\\
\small{$^3$Centro Brasileiro de Pesquisas F\'{\i}sicas,}\\
\small{22290-180 Rio de Janeiro RJ, Brazil.}\\
\small{$^4$Departamento de F\'{\i}sica Te\'orica,}\\
\small{Universidade Estadual do Rio de Janeiro,}\\
\small{20550-013 Rio de Janeiro RJ, Brazil.}}
\maketitle

\begin{abstract}
We analyse the concept of active gravitational mass for
Reissner-Nordstr\"om spacetime in terms of scalar polynomial
invariants and the Karlhede classification. We show that while the
Kretschmann scalar does not produce the expected expression
for the active gravitational mass, both scalar polynomial
invariants formed from the Weyl tensor, and the Cartan scalars, do.
\end{abstract}

\newpage

\section{Introduction}
In a recent paper~\cite{Henry} the Kretschmann scalar is attributed with
the role of characterizing the curvature of spacetime: in the words of
the author, {\it allowing one to ``see'' the black hole} and so {\it any
possible black hole can be visualized realistically}.
Besides this scalar, the scalars of Chern-Pontryagin and Euler
have also been studied
\cite{Cherubini} in a similar context. In this letter we
look at the invariant characterization of the curvature of
spacetime using a different approach. 

We restrict ourselves to spherical symmetry and to the analysis
of the Reissner-Nordstr\"om (RN) spacetime \cite{Reis,Nord}
\begin{equation}
\d s^2=A\,\d t^2-A^{-1}\d r^2-r^2(\d \theta^2+\sin^2\theta\,\d \phi^2), \label{1}
\end{equation}
where $A$ is defined by
\begin{equation}
A\equiv 1-\frac{2M}{r}+\frac{Q^2}{r^2}, \label{2}
\end{equation}
with $M$ the mass parameter and $Q$ the charge.

By taking a spherical surface, $\Sigma$,
centred
at $r=0$ in the coordinate
system describing the
metric (\ref{1}) we expect that the gravitational
field outside $\Sigma$ does not affect the field inside $\Sigma$. The
mass $M$ as well as the charge $Q$ produce gravitational field. Hence
it is reasonable to expect that we can find a quantity at $r$, that
we call gravitational mass $m(r)$, that takes the place of $M$ and $Q$
by producing a corresponding Schwarzschild (S) gravitational field \cite{Schwa} at
$\Sigma$. Our approach to find an eligible invariant expression to
describe physically the curvature should be one that reproduces the
corresponding S expression with $m(r)$.

\section{The active gravitational mass}
In a recent paper~\cite{Barbachoux} different concepts of gravitational
mass for the RN spacetime are discussed. The favoured
definition, by the authors, is the active gravitational mass, $m_a(r)$,
obtained by Whittaker~\cite{Whittaker}. Starting from the active
gravitational mass density, $\mu$,
defined by Whittaker~\cite{Whittaker} and Tolman~\cite{Tolman} as
\begin{equation}
\mu=E^0_0-E^i_i,
\label{3}
\end{equation}
where $E^{\alpha}_{\beta}$ is the electromagnetic energy tensor, the
active gravitational mass inside a volume $V$ is given by
\begin{equation}
m_a(r)=\int_V \mu(-g)^{1/2}\,\d x^1\,\d x^2\,\d x^3,
\label{4}
\end{equation}
where $g$ is the four-dimensional determinant of the metric. Applying~(\ref{4})
to the metric~(\ref{1}) we find
\begin{equation}
m_a(\infty)-m_a(r)=\int^{\infty}_{r}\frac{Q^2}{r^2}\,\d r.
\label{5}
\end{equation}
Since $M$ prevails asymptotically in RN spacetime we assume
$m_a(\infty)=M$, and~(\ref{5}) becomes
\begin{equation}
m_a(r)=M-\frac{Q^2}{r}.
\label{AGM}
\end{equation}

Considering a general anisotropic
charged static fluid source for the
RN spacetime it is possible to prove~\cite{Ponce} that, if the energy
conditions are satisfied, the maximum charge allowed inside the fluid
sphere of boundary radius $r_b$, is
\begin{equation}
Q^2_{max}=Mr_b.
\label{7}
\end{equation}
Another interpretation~\cite{Ponce} of~(\ref{7}), is that the energy
conditions impose a lower limit for the size of the fluid distribution,
\begin{equation}
r_{b_{min}}=\frac{Q^2}{M}.
\label{8}
\end{equation} 
Hence, condition~(\ref{8}) imposes on~(\ref{AGM}) that the active
gravitational mass $m_a\geq 0$.

Some further reasons that suggest that $m_a(r)$ is a plausible definition
are the following.
The equations governing the radial geodesics and the circular geodesics in
the equatorial plane of RN spacetime for a chargeless test particle are,
respectively,
\begin{equation}
\frac{\d ^2r}{\d \tau^2}=-\frac{m_a(r)}{r^2}, \;\; \left(\frac{\d \phi}{\d \tau}\right)^2
=\frac{m_a(r)}{r^3},
\label{9}
\end{equation}
where $\tau$ is the proper time. We see from~(\ref{9}) that locally
$m_a(r)$ casts the
equation of motion
in a Newtonian like form. Another
reason is that at the event horizon of RN spacetime the active
gravitational
mass is equal to the geometrical S mass of the RN
spacetime~\cite{Barbachoux}.  Furthermore, as shown in~\cite{Herrera},
the active gravitational mass allows one to obtain a better grasp of
the physical content of the matter when analysing the energy content of
a slowly collapsing gravitating sphere.

\section{The Weyl and Cartan scalars}
As is well known, the Riemann curvature tensor can be separated
in a coordinate invariant way into the Weyl tensor,
the Ricci tensor and the curvature scalar,
from which we may deduce that the Weyl is generated
only by the gravitational field. From the Weyl tensor we may construct
the {\em Weyl scalar\/}
${\mathcal{C}}\equiv C^{\alpha\beta\gamma\delta}C_{\alpha\beta\gamma\delta}$,
which, for the RN spacetime~(\ref{1}), is
\begin{equation}
{\mathcal{C}}_{\mathrm RN}=\frac{48}{r^6}\left(M-\frac{Q^2}{r}\right)^2.
\label{10}
\end{equation}
Comparing~(\ref{10}) with the corresponding form for the Schwarzschild
spacetime with mass parameter $m$,
\begin{equation}
{\mathcal{C}}_{\mathrm S}=\frac{48}{r^6}\,m^2,
\label{11}
\end{equation}
we see that the expressions are equivalent if we make the
identification~(\ref{AGM}) for $m_a(r)$.

Another way of describing a spacetime invariantly is by its
Karlhede classification~\cite{Karlhede}. Using a basis fixed up to the
isotropy group of the spacetime, the frame components of the decompositions
of the Riemann tensor and its covariant derivatives become invariantly
defined scalars for the spacetime. These scalars have been referred to
in the literature~\cite{Silva} as the Cartan scalars
for the spacetime. The Cartan scalars
provide a more refined invariant characterization of the spacetime than
scalar polynomial invariants since,
for example, they distinguish between the Minkowski spacetime and special
plane wave solutions, for which all scalar polynomial invariants vanish. Notice that although Cartan scalars transform like scalars under
 coordinate transformations, they transform like spinor components under
 basis transformations

In spinor notation, one of the Cartan scalars for the RN spacetime
is the only non-null component of the Weyl spinor
\begin{equation}
(\Psi_2)_{\mathrm RN}=-\frac{1}{r^3}\left(M-\frac{Q^2}{r}\right),
\label{12}
\end{equation}
For the Schwarzschild spacetime with mass parameter $m$
the corresponding Cartan scalar is
\begin{equation}
(\Psi_2)_{\mathrm S} = -\frac{m}{r^3}
\label{13}
\end{equation}
and once again we identify equivalent forms if we use the
definition~(\ref{AGM}) for the gravitational mass.
Several authors \cite{Szekeres}, \cite{Glass} associate $\Psi_2$
with the purely gravitational energy that arises from the Weyl tensor
for a collapsing fluid sphere.

In passing we note that the Kretschmann scalar
${\mathcal{R}}=R^{\alpha\beta\gamma\delta}R_{\alpha\beta\gamma\delta}$
for the RN spacetime is
\begin{equation}
{\mathcal{R}}_{RN}
=\frac{46}{r^6}\left[\left(M-\frac{Q^2}{r}\right)^2+\frac{Q^4}{6r^2}\right],
\label{14}
\end{equation}
which cannot be written in the corresponding Schwarzschild form with $m_a(r)$.
Returning to the beginning of this article, while this scalar may be useful
for measuring curvature near a Schwarzschild black hole, it does not seem
to us to be the case that with it ``any possible black hole can be visualized
realistically''. The Weyl and Cartan scalars seem to be better suited for
this purpose in the non-vacuum cases.

\end{document}